\documentclass[12pt,titlepage]{article}
\usepackage{epsfig}
\usepackage{lscape}
\usepackage{amsmath,epsf,color,colordvi,shadow,pifont}

\usepackage{amsmath}
\usepackage{url}
\usepackage{graphicx}
\usepackage{amssymb}
\usepackage{amsmath,amsfonts,amssymb} 
\usepackage{graphicx} 
\usepackage{lmodern}
\usepackage{ctib} 
\usepackage{arabtex} 
\usepackage{CJKutf8} 
\usepackage{multicol}
\usepackage{hyperref} 

\begin{document}

\begin{titlepage}
\begin{center}
{\LARGE {\bf  Dynamics of the \\ Cosmological Apparent Horizon: Surface Gravity \& Temperature}}
 \\
\vskip 1cm

{\large Alexis Helou\footnote{alexis.helou@apc.univ-paris7.fr}} \\
{\em AstroParticule et Cosmologie, 
Universit\'e Paris Diderot, CNRS, CEA, Observatoire de Paris, Sorbonne Paris 
Cit\'e} \\
{B\^atiment Condorcet, 10, rue Alice Domon et L\'eonie Duquet,\\
F-75205 Paris Cedex 13, France}
\end{center}

\centerline{ {\bf Abstract}} In the context of thermodynamics applied to our cosmological apparent horizon, we explicit in greater details our previous work which established the Friedmann Equations from projection of Hayward's Unified First Law. In particular, we show that the dynamical Hayward-Kodama surface gravity is perfectly well-defined and is suitable for this derivation. We then relate this surface gravity to a physical notion of temperature, and show this has constant, positive sign for any kind of past-inner trapping horizons. Hopefully this will clarify the choice of temperature in a dynamical Friedmann-Lemaître-Roberston-Walker spacetime.

\indent 

\vfill  
\end{titlepage}

\tableofcontents


\section*{Introduction}
Since the seminal work of Jacobson \cite{Jacobson:1995ab} which recovered Einstein Equations from the Clausius Relation applied to a local Rindler horizon, many attempts were made to adapt this idea to a Friedmann-Lemaître-Robertson-Walker spacetime. The first challenge was to find the right horizon on which to work. The local Rindler horizon, where the expansion $\theta$ of null geodesic congruence vanishes, finds its quasi-local equivalent in the cosmological apparent horizon, on which the ingoing expansion $\theta_-$ vanishes \cite{Bak:1999hd}. Then the Unified First Law of Hayward \cite{Hayward:1997jp}, initially designed for black holes, was identified as the relevant relation, along with the tools needed to build it (Misner-Sharp energy \cite{Misner:1964je}, Kodama vector \cite{Kodama:1979vn}). In particular, the associated, dynamical surface gravity was computed. However, no one succeeded in establishing Friedmann Equations from the Unified First Law using this dynamical $\kappa$ (only the reverse way, from gravity to thermodynamics, was achieved \cite{Cai:2006rs}). 

Very recently, we presented a clean and perfectly well-defined way of recovering Friedman from the UFL \cite{Binetruy:2014ela}. Nevertheless, later works continued to claim that the dynamical surface gravity was not relevant, and that the static form $\kappa=1/R_A$ was to be used. 

In the present work, we intend to clarify the situation. In Section \ref{section_dynamical_setup}, we present the tools we need in our dynamical, spherically symmetric framework. In Section \ref{section_Friedmann_Eq}, we reproduce the computation of our previous work \cite{Binetruy:2014ela}, in a more detailed manner. Section \ref{section_misconceptions} compares the logic of our derivation with the logic of the usual, approximate computation that was up to now found in the literature. This will give us some extra information on the way to project the Unified First Law. Section \ref{section_temperature} shows how to relate our surface gravity to a physical notion of temperature, in the tunneling method for modelling Hawking radiation. In the last section, we show that the sign of our temperature does not depend on the causal nature of the horizon, and that a change of sign occurs only when one departs from description of a past-inner horizon. We conclude that the dynamical surface gravity and associated temperature are perfectly well-defined for our expanding Universe, and pose no problem in recovering the Friedmann Equations from the Unified First Law.

\section{The Unified First Law in spherically symmetric spacetimes}
 \label{section_dynamical_setup}
\subsection{Hayward's Unified First Law}
We will work here in the context of spherically symmetric spacetimes:
\begin{equation}
ds^2 = \gamma_{ij}(x)dx^idx^j +R^2(x)d\Omega^2 \ ,
 \label{eq_spherical}
\end{equation}
with signature (-,+,+,+) and coordinates $x^0=t$, $x^1=r$, R being a function of $t$ and $r$.
Within this symmetry, a relevant definition for the energy is the Misner-Sharp one :
\begin{equation}
E(R) \equiv \frac{R}{2G}(1-\nabla^a R \nabla_a R ) \ ,
 \label{eq_Misner_Sharp}
\end{equation}
which is the total gravitational energy inside a sphere of physical radius $R$ \cite{Misner:1964je}. On the apparent/trapping horizon, $\nabla^a R \nabla_a R=0$, and the Misner-Sharp energy reduces to the Schwarzschild mass. 
Another important quantity is the energy-supply covector:
\begin{equation}
\psi_a \equiv T_a^{\phantom{a}b}\nabla_b R + \omega \nabla_a R  \ ,
 \label{eq_energy_supply}
\end{equation}
with an energy-momentum tensor $T_{ab}$, the trace of which defines: 
\begin{equation}
\omega \equiv -\frac{1}{2}T^{ij} \gamma_{ij} \ .
 \label{eq_omega}
\end{equation}
Then the Unified First Law of Thermodynamics established by Hayward \cite{Hayward:1997jp} is:
\begin{equation}
\nabla_a E = A \psi_a + \omega \nabla_a V \ ,
 \label{eq_Unified_First_Law}
\end{equation}
where we use flat area and volume, $A=4\pi R^2 \ $ and $V=\frac{4}{3}\pi R^3$.

\subsection{Definition of the surface gravity for dynamical spacetimes}
In static spacetimes, one may encode the time-translational invariance into a Killing vector field $\xi^a$, which satisfies:
\begin{equation}
\nabla_a \xi_b + \nabla_b \xi_a=0 \ .
 \label{eq_Killing}
\end{equation}
The usual notion of surface gravity $\kappa$ is defined from the Killing:
\begin{equation}
\xi^a(\nabla_a \xi_b - \nabla_b \xi_a)=2\kappa \xi_b \ .
 \label{eq_surface_grav_static}
\end{equation}
However, in a dynamical spacetime, one does not have a time-translational Killing field anymore. Still, this notion may be generalized to the notion of Kodama field $K^a$ \cite{Kodama:1979vn}:
\begin{equation}
K^a \equiv \epsilon^{ab}_{\perp} \nabla_b R \ ,
 \label{eq_Kodama}
\end{equation}
where $\epsilon^{ab}_{\perp}$ is the (1+1) antisymmetric tensor in the (t,r) plane. Unlike the Killing, the Kodama does not give a direction of invariance in time, but it still provides a preferred time-direction. It therefore does not satisfy to the Killing Equation Eq.\eqref{eq_Killing}, but to a slightly altered form of it:
\begin{equation}
K^a(\nabla_a K_b + \nabla_b K_a)= 8\pi G R \psi_b \ .
 \label{eq_Kodama_equation}
\end{equation}
One may then understand the energy-supply as a measure of departure from static symmetry. Both the Misner-Sharp energy and the energy-supply vector may be expressed in terms of the Kodama vector, which we take as a sign of coherency between these quantities.

The defining equation Eq.\eqref{eq_surface_grav_static} for the surface gravity $\kappa$ must also be adapted, since $K^a (\nabla_a K_b - \nabla_b K_a)$ is no longer colinear to $K_b$: 
\begin{equation}
K^a(\nabla_a K_b - \nabla_b K_a)=2\kappa \nabla_b R  \neq 2\kappa K_b \ .
 \label{eq_surface_grav_dyn}
\end{equation}
We thus have a perfectly well-defined notion of surface gravity. It may be expressed as:
\begin{align}
&\kappa = \frac{1}{2\sqrt{-\gamma}}\partial_i[\sqrt{-\gamma}\gamma^{ij}\nabla_j R] 
\label{eq_surface_grav1} \\
\text{or } \nonumber \\ 
&\kappa = G\frac{E}{R^2} - 4\pi \omega R G \ .
 \label{eq_surface_grav2}
\end{align}
This is sometimes referred to as the ``Hayward-Kodama'' surface gravity in the literature.

\subsection{Reformulation of the Unified First Law}
Starting from Eq.\eqref{eq_Unified_First_Law}, we may now rewrite the Unified First Law as:
\begin{align}
A\psi_a &= \nabla_a E  - \omega \nabla_a V \nonumber \\
        &= R \nabla_a \left(\frac{E}{R}\right) + \frac{E}{R} \nabla_a (R) -\frac{\omega}{3}(A\nabla_a R+R\nabla_a A) \nonumber \\
        &= R \nabla_a \left(\frac{E}{R}\right) + \frac{E}{8\pi R^2} \nabla_a A -\frac{\omega}{3}\left(\frac{A}{8\pi R}\nabla_a A +R\nabla_a A\right) \nonumber \\
        &= R \nabla_a \left(\frac{E}{R}\right) + \nabla_a A \left( \frac{E}{8\pi R^2} -\frac{\omega}{3}\frac{A}{8\pi R} -\frac{\omega R}{3}\right) \nonumber \\
        &= R \nabla_a \left(\frac{E}{R}\right) + \nabla_a A \left( \frac{E}{8\pi R^2} -\frac{\omega R}{6} -\frac{\omega R}{3}\right) \nonumber \\
        &= R \nabla_a \left(\frac{E}{R}\right) + \frac{1}{8\pi G} \nabla_a A \left( G\frac{E}{R^2} -4\pi \omega R G \right) \nonumber \\
        &= R \nabla_a \left(\frac{E}{R}\right) +  \frac{\kappa}{8\pi G} \nabla_a A
 \label{eq_unified_first_law_reform}
\end{align}
where in the last equality we recognized the surface gravity $\kappa$ from Eq.\eqref{eq_surface_grav2}.
Here one should not get confused: this is the full Unified First Law, not just an independent expression of the energy-supply term. The fact that the Unified First Law is written the other way around should not occult that Eq.\eqref{eq_unified_first_law_reform} is the full law.

\section{Recovering Friedmann Equations from the Unified First Law projected on the Apparent Horizon}
 \label{section_Friedmann_Eq}
In this section, we give a detailed explanation of our derivation of the Friedmann Equations. The original computation may be found in \cite{Binetruy:2014ela}.
\subsection{The Unified First Law projected}
Let us now work in a homogeneous and isotropic Friedmann-Lemaître-Robertson-Walker (FLRW) Universe:
\begin{equation}
ds^2 = -dt^2 + \frac{a(t)^2}{1-k r^2}dr^2 + R^2 d\Omega^2  \ , \\\ \text{   } \\\  R=a(t)r \ .
 \label{eq_FLRW}
\end{equation}
This spacetime has an apparent/trapping horizon satisfying $\nabla^a R \nabla_a R=0$, and so its radius is:
\begin{equation}
R_A = \left( H^2 + \frac{k}{a^2} \right)^{-1/2} \ ,
 \label{eq_apparent_radius}
\end{equation}
which is nothing else than the Hubble radius $H^{-1}$ in the flat case ($k=0$). We also give the time-derivative of the apparent horizon radius:
\begin{equation}
\dot{R}_A = -H R_A^3 \left( \dot{H} - \frac{k}{a^2} \right) \ .
 \label{eq_apparent_radius_deriv}
\end{equation}
We have shown in Eq.\eqref{eq_unified_first_law_reform} that the Unified First Law writes:
\begin{equation}
A\psi_a = \frac{\kappa}{8\pi G}\nabla_a A + R \nabla_a \left(\frac{E}{R}\right) \ ,
 \label{eq_Apsi}
\end{equation}
where the Misner-Sharp energy is, in this cosmological context:
\begin{equation}
E=\frac{R^3}{2G}\left(H^2+\frac{k}{a^2}\right)=\frac{R^3}{2GR_A^2} \ ,
 \label{eq_energy}
\end{equation}
so that the last term in Eq.\eqref{eq_Apsi} reads:
\begin{equation}
R \nabla_a \left(\frac{E}{R}\right)= \frac{R}{2G} \nabla_a \left(\frac{R^2}{R_A^2}\right) \ .
 \label{eq_vanishing_term}
\end{equation}
It is often stated in the literature that this term vanishes on the horizon. This is completely false, except if one means that it vanishes ``along'' the horizon. The quantity in the covariant derivative is indeed constant when evaluated at $R=R_A$, but this is also true for any other fixed value of $R$. Therefore, if one wants to dismiss this term in order to identify a Clausius relation in $A\psi=\frac{\kappa}{8\pi G}dA$, one should keep in mind that it implies all the following computations hold only along a surface of constant $R$. In other words, in order to get rid of the term in question, one must project the Unified First Law Eq.\eqref{eq_Apsi} on a vector $t^a$ which is tangent to the horizon. Since the apparent horizon is defined as the surface where $\nabla^a R \nabla_a R=0$, this tangent vector is defined by  $t^a \nabla_a [\nabla^b R \nabla_b R]=0$. For the FLRW Universe, it yields:
\begin{equation}
t^a = \frac{1}{2}\left( 1 , -\frac{RH}{a}\left(1-\frac{\dot{R}_A}{H R_A}\right) , 0, 0 \right) \ .
 \label{eq_tangent_vector}
\end{equation}
The Unified First Law may be identified to the Clausius Relation only once it is projected on this tangent vector $t^a$. It then reads:
\begin{equation}
t^a (A \psi_a) =  t^a \left(\frac{\kappa}{8\pi G}\nabla_a A  + R \nabla_a \left(\frac{E}{R}\right) \right)= t^a \left(\frac{\kappa}{8\pi G}\nabla_a A \right)  \ .
 \label{eq_Clausius}
\end{equation}
Let us now detail the computations for each of the three terms in this projected law.

\subsection{The Friedmann Equation}
 \label{subsection_Friedmann}
\subsubsection*{The vanishing term}
We here check that the second term on the right-hand side of  Eq.\eqref{eq_Clausius} does vanish when projected onto $t^a$:
\begin{align}
t^a \nabla_a \left(\frac{E}{R}\right) &= \frac{t^t}{2G}\partial_t \left( \frac{R^2}{R_A^2}\right) +\frac{t^r}{2G}\partial_r \left( \frac{R^2}{R_A^2}\right) \nonumber \\
        &= \frac{1}{4G} r^2 \partial_t \left( \frac{a^2}{R_A^2} \right) - \frac{1}{4G} \frac{RH}{a}\left(1-\frac{\dot{R}_A}{H R_A}\right) \frac{a^2}{R_A^2} \partial_r (r^2)  \nonumber \\
        &= \frac{1}{4G} r^2  \left( \frac{2a\dot{a}R_A^2 - a^2 2 R_A \dot{R}_A}{R_A^4} \right) - \frac{1}{2G} \frac{R^2 H}{R_A^2}\left(1-\frac{\dot{R}_A}{H R_A}\right)    \nonumber \\
        &= \frac{1}{2G}\frac{R^2 H}{R_A^2}   -\frac{1}{2G} \frac{R^2\dot{R}_A}{R_A^3} - \frac{1}{2G} \frac{R^2 H}{R_A^2} +  \frac{1}{2G} \frac{R^2\dot{R}_A}{R_A^3} \nonumber \\
        &= 0         \ .
 \label{eq_vanishing_term}
\end{align}
Note that this is true for any radius $R$, not just for $R=R_A$.

\subsubsection*{The energy-supply term}
Now let us compute the non-vanishing terms in Eq.\eqref{eq_Clausius}. Assuming the energy-momentum of a perfect fluid in our FLRW Universe, one gets the energy-supply covector from Eq.\eqref{eq_energy_supply} :
\begin{equation}
\psi_a = \frac{1}{2}(\rho +p) \left( -RH , a , 0, 0\right) \ .
 \label{eq_energy_supply_FLRW}
\end{equation}
Its projection is then:
\begin{align}
t^a (A \psi_a) &= - \frac{1}{4}ARH(\rho +p)  - \frac{a}{4}(\rho +p)\frac{ARH}{a}\left(1-\frac{\dot{R}_A}{H R_A}\right)\nonumber \\
 			   &= - \frac{1}{2}ARH(\rho +p)  + (\rho +p)\frac{AR}{4}\frac{\dot{R}_A}{R_A}\nonumber \\
               &= - 2\pi HR^3(\rho +p)   \left(1- \frac{\dot{R}_A}{2HR_A} \right)   \ .   
 \label{eq_psi_projected}
\end{align}

\subsubsection*{The area term}
The right-hand side of Eq.\eqref{eq_Clausius} reads:
\begin{align}
t^a \left(\frac{\kappa}{8\pi G}\nabla_a A \right) &= \frac{\kappa}{4G} \left( \partial_t (R^2)  -\frac{RH}{a}\left(1-\frac{\dot{R}_A}{H R_A}\right) \partial_r(R^2) \right)   \nonumber \\
	 &= \frac{\kappa}{4G} \left( 2HR^2 - 2HR^2\left(1-\frac{\dot{R}_A}{H R_A}\right) \right) \nonumber \\
	 &= \frac{\kappa R^2}{2G}\frac{\dot{R}_A}{R_A}  \ .
 \label{eq_dA_projected}
\end{align}
In a FLRW spacetime, the surface gravity Eq.\eqref{eq_surface_grav1} is:
\begin{align}
\kappa &= -\frac{R}{2} \left( 2H^2 +\dot{H} +\frac{k}{a^2}\right)  \nonumber  \\
	   &= -\frac{R}{2} \left( \frac{2}{R_A^2} -\frac{\dot{R}_A}{H R_A^3}\right)  \nonumber  \\
	   &= -\frac{R}{R_A^2} \left( 1 -\frac{\dot{R}_A}{2HR_A}\right) \ ,
 \label{eq_surfgravFLRW}
\end{align}
so that Eq.\eqref{eq_dA_projected} also writes:
\begin{equation}
t^a \left(\frac{\kappa}{8\pi G}\nabla_a A \right)= \frac{\kappa R^2}{2G}\frac{\dot{R}_A}{R_A} = -\frac{R^3 }{2G}\frac{\dot{R}_A}{R_A^3}\left( 1 -\frac{\dot{R}_A}{2HR_A}\right)  \ .
 \label{eq_dA_projected_final}
\end{equation}

\subsubsection*{The Friedmann Equation}
Finally, using the results of our projections Eqs. \eqref{eq_vanishing_term}, \eqref{eq_psi_projected} and \eqref{eq_dA_projected_final} into the full projected Unified First Law Eq.\eqref{eq_Clausius}, we get:
\begin{align}
t^a (A \psi_a) &= t^a(\frac{\kappa}{8\pi G}\nabla_a A ) \nonumber \\
- 2\pi HR^3(\rho +p)   \left(1- \frac{\dot{R}_A}{2HR_A} \right)   &= -\frac{R^3 }{2G}\frac{\dot{R}_A}{R_A^3}\left( 1 -\frac{\dot{R}_A}{2HR_A}\right)  \nonumber \\
2\pi H(\rho +p)     &= \frac{1}{2G}\frac{\dot{R}_A}{R_A^3} \nonumber \\
2\pi H(\rho +p)     &= -\frac{1}{2G} H\left( \dot{H}-\frac{k}{a^2} \right)\nonumber \\
-4\pi G(\rho +p)    &= \dot{H}-\frac{k}{a^2}  \ .
 \label{eq_Friedmann_eq}
\end{align}
This is indeed the second Friedmann Equation.
Note that up until now, we have not made any mention of the notions of temperature or entropy. We \emph{do not need} to assume any such relations as $T=\frac{\kappa}{2\pi}$ or $S=\frac{A}{4G}$ to derive the Friedmann Equation from the Unified First Law. The introduction of temperature and entropy should come afterwards, when one tries to identify the projected Unified First Law, $t^a (A \psi_a) = t^a(\frac{\kappa}{8\pi G}\nabla_a A )$, with a Clausius relation of the form $\delta Q = TdS$. However, one should be extremely careful with the \emph{physical interpretation} of these quantities. Defining a new quantity $T$ as $T=\frac{\kappa}{2\pi}$ does not mean it is a temperature!

 We will postpone to Section \ref{section_temperature} the interpretation of our surface gravity $\kappa$ in terms of temperature. But first we deem it useful to present the common way of recovering Friedmann Equations from thermodynamics, the conceptual differences it has with the above derivation, and the extra-information we can get from it.

\section{The original derivation and a new way of projecting the Unified First Law}
 \label{section_misconceptions}

\subsection{Comparison with the original derivation}
The original computations were going from Thermodynamics to Friedmann Equations \cite{Cai:2005ra}, and then the other way around \cite{Cai:2006rs}. Many other works followed these seminal computations, and applied them to several theories of gravity, but with no major variation of the line of reasoning. The logic of all these derivations is quite different from ours, as well as the needed assumptions. The original derivation \cite{Cai:2005ra} starts from an infinitesimal time variation of the Unified First Law expressed in the $(t,R,\theta,\phi)$ coordinates instead of the previous $(t,r,\theta,\phi)$. Then one invokes the first law of thermodynamics $-dE=TdS$, with a temperature $T$ assumed to have the usual (but static!) form $T=1/2\pi R_A$ and an entropy $S$ that is assumed to scale as the area of the apparent horizon. (Note once again that the derivation detailed in Section \ref{section_Friedmann_Eq} is completely free of such assumptions). One still has to evaluate this equation on the apparent horizon to finally get the Friedmann Equation (which makes it valid only at the horizon).

Therefore the reasoning goes as follows: on the one hand, the time variation of the Unified First Law gives an expression for $dE$. On the other hand, one must call for the usual first law of thermodynamics (the Clausius Relation) to express $dE$ in another way. Equating both expressions for $dE$ yields the Friedmann Equation. This is drastically different from the reasoning in the computation of Section \ref{section_Friedmann_Eq}, which only uses the Unified First Law and makes no assumption on temperature and entropy. However it is true that our computation, though it does not use the Clausius Relation, relies on another hypothesis: that the energy takes the Misner-Sharp form Eq.\eqref{eq_energy}. We see this hypothesis as better motivated than those on temperature and entropy. Hayward has justified the use of the Misner-Sharp energy in spherical symmetry \cite{Hayward:1994bu}, and this definition is widely used by the community. The logic of the two different derivations can be sketched as follows:
\begin{itemize}
\item our computation: Unified First Law + Misner-Sharp = Friedmann (which may also be interpreted \emph{a posteriori} as a Clausius Relation, with temperature and entropy), 
\item common computation: Unified First Law + Clausius Relation (with temperature and entropy)= Friedmann. 
\end{itemize} 
Thus, one can wonder whether the two situations are equivalent: the common derivation may take for hypothesis the conclusions of our derivation, and vice-versa. If it were the case, then the common computation should recover the Misner-Sharp energy as a conclusion. However, integrating the infinitesimal $dE=-TdS=-\frac{\dot{R}_A}{G}dt$, one gets:
\begin{equation}
E = -\frac{R_A}{G}  \ ,
 \label{eq_E_Cai}
\end{equation}
which is a strange form for the energy, and not the Misner-Sharp one. We take this as a hint that the original computation holds only as an approximation. Indeed, researchers of the field who tried to justify the form of the temperature $T=1/2\pi R_A$, started from the full Kodama-Hayward surface gravity Eq.\eqref{eq_surfgravFLRW} evaluated on the horizon, and then argued that the process should be adiabatic and the time derivative $\dot{R}_A$ negligible. In this sense, $T=1/2\pi R_A$ can only be an approximate temperature at best. Section \ref{section_Friedmann_Eq} does not use this approximation.

Therefore the exact derivation of Friedmann Equations from thermodynamics of the horizon was not achieved, until a rigorous computation was provided in \cite{Binetruy:2014ela}. However, the Cai-Kim derivation \cite{Cai:2005ra} turns out to be very instructive when performed in a projective manner: it yields another vector field on which to project the UFL in order to recover Friedmann Equations.

\subsection{Projection of the Unified First Law on $\partial_t$}
 \label{section_cleaner_Cai}
Indeed, when considering only a time variation in the Unified First Law, one is doing nothing else than projecting onto a vector $t'^a$ which is just $\partial_t$:
\begin{equation}
t'^a = (1,0,0,0)_{(t,R,\theta,\phi)} \ .
 \label{eq_time_vector}
\end{equation}
The Unified First Law reads, still in the $(t,R,\theta,\phi)$ coordinates:
\begin{align}
A t'^a \psi_a &= R t'^a \nabla_a \left(\frac{E}{R}\right) +  \frac{\kappa}{8\pi G} t'^a \nabla_a A \nonumber \\
-AHR(\rho +p) &= R  \partial_t \left(\frac{R^2}{2GR_A^2}\right) +  \frac{\kappa}{8\pi G} 8\pi R \nabla_t R \nonumber \\
-4\pi HR^3(\rho +p) &=  \frac{R^3}{2G} \partial_t \left( H^2 +\frac{k}{a^2}\right) +  0 \nonumber \\
-4\pi HR^3(\rho +p) &=  \frac{HR^3}{G} \left(  \dot{H} -\frac{k}{a^2}\right) 
 \label{eq_Cai_neat}
\end{align}
The Friedmann Equation is recovered straightforwardly, once again without any assumptions on entropy or temperature (though assuming a Misner-Sharp energy). We have just found another vector field on which to project the Unified First Law! An interesting point is that this new vector $t'^a$ turns out to be colinear to the Kodama vector, which reads in the $(t,R,\theta,\phi)$ coordinates:
\begin{equation}
 K^a = \sqrt{1-k\frac{R^2}{a^2}} \times \left( 1,0,0,0 \right) =  \sqrt{1-k\frac{R^2}{a^2}} \times t'^a \ .
\end{equation}
Thus projecting on the Kodama also yields the Friedmann Equations! This is not so surprising, as $K^a$ is not \emph{any} vector: it indeed gives a preferred time-direction.

A puzzling difference arises when one compares this new projection to that of Section \ref{section_Friedmann_Eq}. There it is shown that, when projecting on $t^a$, the vanishing term and the term yielding $\left(  \dot{H} -\frac{k}{a^2}\right)$ are the $\nabla_a \left(\frac{E}{R}\right)$ term and the $(\frac{\kappa}{8\pi G}\nabla_a A) $ term respectively. It is exactly the contrary for projection on $K^a$. How is one to identify a Clausius relation $dE=TdS$ in the projected Unified First Law $A K^a \psi_a = R K^a \nabla_a \left(\frac{E}{R}\right)$? What are the temperature and entropy now? It is not possible to identify the surface gravity $\kappa$ of Eq.\eqref{eq_surfgravFLRW} in the previous equation. Moreover, although it was identified in the first projection (on $t^a$), $\kappa$ appeared on both sides of the Unified First Law, and was thus of no use for the establishment of the Friedmann law. This seems to point at the Clausius relation as not being fundamental in the derivation of Friedmann Equations. In the end, the original derivation, written in this projective manner, is the one that does not even need to express the surface gravity $\kappa$ and that demands no assumption on the temperature. 
We summarise the situation in Table \ref{table_projection}.

\begin{table}
\begin{tabular}{|l|c|c|c|}
  \hline
  projection & $A\psi_a$ & $\frac{\kappa}{8\pi G}\nabla_a A$ & $R\nabla_a \left(\frac{E}{R}\right)$  \\
  \hline
  on tangent field $t^a$ & $2\pi \kappa H R^2 R_A^2(\rho +p)$ & $\frac{\kappa}{2G} \frac{R^2 \dot{R}_A}{R_A}$  & 0\\
  \hline
  on Kodama field $K^a$ & $-4\pi HR^3(\rho +p)$ & 0 & $-\frac{R^3}{G R_A^3}\dot{R}_A$\\
  \hline
\end{tabular}
\caption{Projection of the UFL on two different vectors, term by term.}
 \label{table_projection}
\end{table}

In the previous sections, we have established that the dynamical surface gravity Eq.\eqref{eq_surface_grav1} poses no problem in the derivation of Friedmann Equations from the Unified First Law. In a spherically symmetric, dynamical framework, it is the relevant quantity to use, along with the Kodama vector and Misner-Sharp energy. In the next section, we will try to see whether this surface gravity can be used to define a notion of temperature.

\section{From Surface Gravity to Temperature}
 \label{section_temperature} 
 
In order to understand whether the surface gravity $\kappa_{\cal H}$ may be related to a notion of temperature, let us use the tunneling approach by Parikh and Wilczek \cite{Parikh:1999mf}, in the Hamilton-Jacobi formulation (see \cite{Hayward:2008jq} for the black hole case).

\begin{center}
\begin{itemize}
\item Past Horizon (FLRW) $\rightarrow$ Retarded Eddington-Finkelstein coordinates:
 \begin{equation}
 ds^2 = -e^{2\Psi}Cdx_-^2 - 2e^{\Psi}dx_-dR  + R^2d\Omega^2 \ . 
 \end{equation}
with $x_-=\eta-\chi$  (conformal time and comoving distance). 
In order to give an idea of what the functions $\Psi$ and $C$ are, we give their expression for a FLRW Universe:
\begin{equation}
e^{\Psi}=\frac{a}{\sqrt{1-kr^2}+RH} \ ,
\end{equation}
\begin{equation}
e^{2\Psi}C=a^2\frac{\sqrt{1-kr^2}-RH}{\sqrt{1-kr^2}+RH} \ , 
\end{equation}
\begin{equation}
C=1-kr^2-R^2H^2 \ ,
 \label{eq_def_C_cosmo}
\end{equation}
but we do not need to specialize to FLRW now, and so we go back to the general case.
\item Kodama vector: $K^a=(e^{-\Psi} ; 0 ; 0 ; 0) \ .$
\item Surface gravity: $\kappa_{\cal H} = \frac{1}{2} \left. \partial_R C\right|_{\cal H} \leqslant 0 \ .$
\item BKW approximation of tunneling probability for a massless scalar field $\phi=\phi_0 \exp(i\mathcal{I})$:
\begin{equation}
\Gamma \propto exp\left(-2\frac{Im\mathcal{I}}{\hbar}\right) \ .
\end{equation}
\item Equation of Motion is Hamilton-Jacobi Equation:
\begin{equation}
g^{ab} \nabla_a \mathcal{I} \nabla_b \mathcal{I}= 0 \ .
\end{equation}
\item Action: $\mathcal{I}=\int\partial_{x_-}(\mathcal{I})dx_- + \int\partial_{R}(\mathcal{I})dR\ .$
\item Kodama energy\footnote{This is a generalization of the Killing energy. Not to be confused with $\omega$ in the Unified First Law.}: $\omega=-K^a\nabla_a \mathcal{I} = -e^{-\Psi}\partial_{x_-}(\mathcal{I}) \ .$ 
\item Wave number: $k=\partial_{R} \mathcal{I} \ .$
\item EoM: $k(Ck+2\omega) = 0 \ .$
 \\$\rightarrow$ $k=0$ (outgoing solution).
 \\$\rightarrow$ $k=-2\omega/C$ (ingoing).
\item Ingoing solution has a pole and contributes to imaginary part of the action: 
 \begin{align}
  Im\mathcal{I} =& Im \int \partial_R (\mathcal{I}) dR \nonumber = Im \int k dR \\ \nonumber
  =& Im \int -\frac{2\omega}{C}dR  \\ \nonumber
  =& Im \int -\frac{\omega}{\kappa_{\cal H}(R-R_A)}dR \ , \nonumber
 \end{align}
since near the horizon, $C \approx \partial_R C (R-R_A)= 2\kappa_{\cal H}(R-R_A)$.
Using Feynman's $i\epsilon$-prescription, we circumvent the real pole from below:
 \begin{align}
  Im &\int -\frac{\omega}{\kappa_{\cal H}(R-R_A-i\epsilon)}dR \nonumber \\ \nonumber
  &= Im \int -\frac{\omega(R-R_A+i\epsilon)}{\kappa_{\cal H}(R-R_A-i\epsilon)(R-R_A+i\epsilon)}dR \\ \nonumber
  &= \int -\frac{\omega\epsilon}{\kappa_{\cal H}(R-R_A-i\epsilon)(R-R_A+i\epsilon)}dR \\ \nonumber
  &=  2i\pi \times \lim_{R\rightarrow R_A+i\epsilon} \left( -\frac{\omega\epsilon(R-R_A-i\epsilon)}{\kappa_{\cal H}(R-R_A-i\epsilon)(R-R_A+i\epsilon)} \right) \\ \nonumber
  &=-\frac{2i\pi \epsilon \omega}{2i \epsilon \kappa_{\cal H}} \\ \nonumber
  &=-\frac{\pi\omega}{\kappa_{\cal H}}\ , 
 \end{align}
where we have used the residue theorem to compute the integral.
\item The tunneling probability takes a thermal form:
\begin{equation}
 \Gamma \propto exp\left(-2Im\mathcal{I}\right) \propto exp(-\omega/T) \Leftrightarrow T= \frac{\omega}{2 Im \mathcal{I}} \ ,
\end{equation}
\item with temperature:
 \begin{equation}
 T=-\frac{\kappa_{\cal H}}{2\pi} \geqslant 0 \ .
 \label{eq_temperature_sign}
\end{equation} 
\end{itemize}
\end{center} 

We indeed get a positive temperature for the thermal spectrum of ingoing modes, \emph{i.e.} of particles tunneling from beyond the horizon to the interior (towards the central observer), as was originally suggested in \cite{Gibbons:1977mu}. The inner character of the horizon for an expanding Universe yields a negative surface gravity, while its past nature imposes a minus sign in the relation between $T$ and $\kappa_{\cal H}$. This combining sign effect is explained in details in \cite{Helou}.

\section{Sign of surface gravity and causal nature of the horizon}
 \label{section_causal_nature} 
\begin{center}
\begin{figure}[h] 
\centering
  \includegraphics[width=15cm,angle=0]{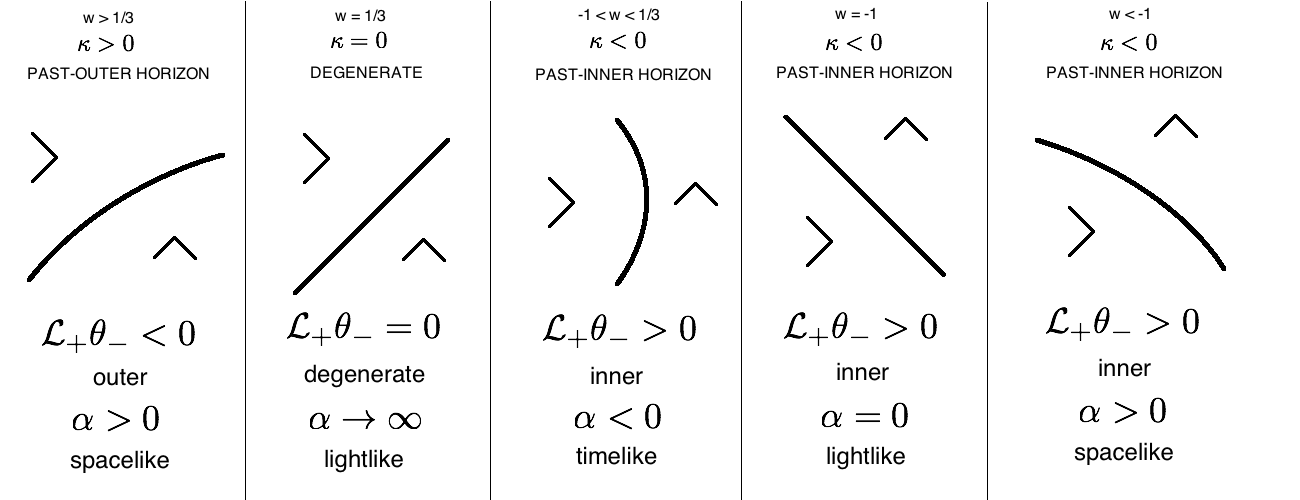}    
   \caption{Causal nature of the horizon and the sign of $\kappa \ .$}
    \label{fig_causal_nature_horizon}
\end{figure}
\end{center}

In a FLRW Universe, we have seen that the surface gravity can be expressed as in Eq.\eqref{eq_surfgravFLRW}. Using Friedmann Equations, one can also express it in terms of a parameter of state $w$: 
\begin{align}
\kappa &= -\frac{R}{2} \left( 2H^2 +\dot{H} +\frac{k}{a^2} \right)  \nonumber  \\
	   &= -\frac{R}{2} \left( 2\left(H^2+\frac{k}{a^2}\right)  +\dot{H} -\frac{k}{a^2} \right)  \nonumber \\
	   &= -\frac{R}{2} \left( \frac{16\pi G}{3}\rho -4\pi G(\rho+p) \right) \nonumber \\
	   &= -\frac{R}{2} \left( \frac{4\pi G}{3}\rho -4\pi G w\rho \right)\nonumber \\
	   &= - 2\pi G R \rho \left( \frac{1}{3} - w \right) \ .
 \label{eq_surfgrav_state_param}
\end{align}
From the above expression for the surface gravity, one can see that it changes sign depending on the energy budget of the FLRW Universe: $\kappa$ is negative for $w<1/3$, positive for $w>1/3$. However, the standard model of cosmology, or $\Lambda$-CDM, describes the history of our Universe by successive eras: inflation, when $w\sim -1$, radiation-domination, $w=1/3$, matter-domination, $w=0$, and the current phase of cosmic acceleration or ``dark energy'', $w\sim -1$. All these phases have $w\leqslant 1/3$, and thus $\kappa$ negative or null. But the signature of the horizon is not constrained: it can be timelike, null or spacelike (see the four cases to the right of Figure \ref{fig_causal_nature_horizon} \footnote{In this figure we have used the Bousso wedge convention, which represents, out of the four null directions of the lightcones, only those along which geodesic congruences are contracting ($\theta<0$). For a Minkowski-like region, these directions are the future and past inner ones (Bousso wedge: $>$). For the trapped region of an expanding cosmology, they are the two past directions  (Bousso wedge: $\wedge$).}). 
The last case to the right of Figure \ref{fig_causal_nature_horizon} represents a spacelike past-inner horizon, with $w<-1$, which could represent a phase of even more accelerated expansion, driven by a hypothetical ``phantom energy''. The previous case to the left is a null past-inner horizon for $w=-1$, and is the usual half of the Penrose diagram for de Sitter Universe. The previous case shows a timelike past-inner horizon for a matter-dominated Universe with $w=0$. The previous drawing (second from the left) is a null past-inner horizon for a radiation-dominated Universe, $w=1/3$. All four cases are past-inner horizons (the last one being degenerate), with $\kappa\leqslant 0$, though the signature can be timelike/lightlike/spacelike.

Only when we tilt the horizon further still do we get a positive $\kappa$ (see the left case on Figure \ref{fig_causal_nature_horizon}). That is only because we have crossed the limit between past-inner horizons ($\mathcal{L}_+ \theta_->0$) and past-outer ones ($\mathcal{L}_+ \theta_-<0$) \footnote{Here the +/- indices refer to outgoing/ingoing future null directions respectively.}. We are no longer describing the same system, but rather something resembling a white-hole \footnote{Nevertheless, past-outer configurations could be used to describe some special expanding cosmologies, for example a stiff-matter cosmology, for $w=1$.}.

Therefore the sign of $\kappa$ does not depend on the signature of the horizon. It is solely dictated by the inner/outer nature of the apparent horizon. The sign of the temperature, however, is a result of both the sign of $\kappa$ and the sign entering in the definition of $T$, as in Eq.\eqref{eq_temperature_sign}. For further details on this combining sign effect, see \cite{Helou}.

Note that this is all exactly similar to the black hole case: a future-outer horizon can be either timelike, null, or spacelike, its surface gravity $\kappa$ will still be of constant, positive sign (the equivalent of the three cases to the right of Figure \ref{fig_causal_nature_horizon}). Of course, if one pushes the horizon further still, the surface gravity will change sign, but that is just because the configuration will have become future-\emph{inner}. We will not be describing a black hole anymore.

A practical way of knowing the causal nature of a horizon is the sign of parameter $\alpha$  \cite{Dreyer:2002mx}:
\begin{equation}
\alpha = \frac{ \mathcal{L}_- \theta_- }{ \mathcal{L}_+ \theta_- }   \ .
 \label{eq_alpha}
\end{equation}
$\alpha$ is negative/null/positive for timelike/lightlike/spacelike horizons respectively. This is true for both inner \emph{and} outer past-horizons.

\section{Conclusions}

We have seen that the sign of the Hayward-Kodama surface gravity \emph{does} depend on the inner/outer character of the trapping horizon. However, the sign of the temperature depends on both the inner/outer \emph{and} the past/future character of the horizon. This is clearly established in \cite{Helou}. The statement is valid for black holes, white holes, expanding cosmologies as well as contracting ones \footnote{It has been stated in \cite{Tian:2014sca} that in the case of a contracting cosmology ($H<0$), one will have to deal with an outer horizon. However in the language of Hayward that we use here, a contracting Universe has a future-inner trapping horizon.}

We have shown in Section \ref{section_dynamical_setup} that the Hayward-Kodama surface gravity is perfectly well-defined in a dynamical context, using the Kodama field (the generalization of the static Killing field). Researchers of the field who have used this definition of surface gravity, have unfortunately let go of the time-dependent part, and approximated $\kappa$ to the usual, static surface gravity $1/2\pi R_A \ .$ Later on, the full Hayward-Kodama surface gravity was rehabilitated \cite{Cai:2006rs}, in a computation going from gravity to thermodynamics. A full, exact computation establishing Friedmann Equations from the Unified First Law, using this dynamical $\kappa$, was nevertheless still needed. It was recently provided in \cite{Binetruy:2014ela}. This derivation was reproduced here in Section \ref{section_Friedmann_Eq}. 

Moreover, the link of our surface gravity to a temperature was also established in \cite{Binetruy:2014ela}, and detailed in this present work, Section \ref{section_temperature}. This quantity $T$ turns out to be positive, and interpretable as the temperature of a thermal spectrum for particles tunneling from outside the horizon towards the central observer. Here, no approximation, truncation, or convenient absolute value have been applied on the temperature. We hope that the present work, added to our previous article \cite{Binetruy:2014ela}, will rehabilitate the Hayward-Kodama surface gravity as the relevant quantity to use in spherically-symmetric dynamical spacetimes.

\textbf{Acknowledgements:} I thank Florian Gautier and Jean-Philippe Bruneton for useful discussions. I thank Pierre Binétruy for comments on the manuscript.

\newpage
\section*{Appendix A}
It is also possible to get the Friedmann Equations from each components of the Unified First Law expressed in the $(t,R,\theta,\phi)$-coordinates. Here again, we use the Misner-Sharp energy.
The $R$-component of the Unified First Law reads:
\begin{align}
\nabla_R{E} = A\psi_R + \omega \nabla_R V &= A\frac{\rho+p}{2} + \frac{\rho-p}{2}\nabla_R V   \nonumber \\		\frac{3}{2G}R^2(H^2+k/a^2) &= 4\pi R^2 \rho \ ,
 \label{eq_unified_first_law_R}
\end{align}
which immediately yields the First Friedmann Equation:
\begin{equation}
H^2+\frac{k}{a^2} =  \frac{8\pi G}{3} \rho \ .
 \label{eq_first_friedmann_eq_from_unified_law}
\end{equation}
Now for the $t$-component:
\begin{align}
\nabla_t{E}= A\psi_t + \omega \nabla_t V &= -HRA(\rho+p) + \frac{\rho-p}{2}\nabla_t V \nonumber \\
 -\frac{R^3 \dot{R}_A}{G R_A^3} &= -4\pi HR^3(\rho+p) \ ,
 \label{eq_unified_first_law_t}
\end{align}
which yields the Second Equation:
\begin{equation}
\dot{H}-\frac{k}{a^2} = -4\pi G (\rho+p) \ .
 \label{eq_second_friedmann_from_unified_law2}
\end{equation}
This is nothing else than the computation of Section \ref{section_cleaner_Cai} (projecting on Kodama amounts to selecting the $t$-component in these coordinates).
Note here that this is once again valid for all $R$, not only on the apparent horizon $R=R_A$.

\underline{Nota Bene:}
\begin{itemize}
\item The UFL in the form of Eq.\eqref{eq_Unified_First_Law}, written in the $(t,R)$ coordinates, yields the First and Second Friedmann Equations directly from its $R$ and $t$-components respectively.
\item The UFL in the form of Eq.\eqref{eq_Unified_First_Law}, written in the $(t,r)$ coordinates, yields the First Friedmann Equation directly from its $r$-component. The $t$-component supplemented with the First Friedmann Equation yields the Second Friedmann Equation.
\item The UFL in the form of Eq.\eqref{eq_unified_first_law_reform}, written in the $(t,R)$ coordinates, yields the Second Friedmann Equation directly from its $t$-component. The $R$-component gives nothing...
\item The UFL in the form of Eq.\eqref{eq_unified_first_law_reform}, written in the $(t,r)$ coordinates, yields the Second Friedmann Equation directly from its $t$-component. The $r$-component gives nothing...
\end{itemize}

The last two remarks are understandable: Eq.\eqref{eq_Unified_First_Law} is not symmetric in $p$ and $\rho$, which allows us to single out an expression for $\rho$ alone. On the other hand, Eq.\eqref{eq_unified_first_law_reform} is symmetric in $p$ and $\rho$, and cannot recover the First Friedmann Equation.

Let us interpret the first remark: the $R$-component of Eq.\eqref{eq_Unified_First_Law} gives the First Friedmann Equation, which links the energy density $\rho$ to the size of the apparent horizon $R_A$. Nowadays, this is dominated by dark energy, $\Omega_{\Lambda}=0.7$, which we take as vacuum energy. An explanation to this has been provided in \cite{Binetruy:2012kx}, in terms of collapse of the quantum wave-function on the most probable state (with highest entropy). Now the $t$-component of Eq.\eqref{eq_Unified_First_Law} gives the Second Friedmann Equation, which links the sum $(\rho+p)$ to the time variation of the apparent horizon $\dot{R}_A$. This, on the contrary, receives no contribution from dark energy: $(\rho_{\Lambda}+p_{\Lambda})=0$. Matter, and radiation, are the contributing components here. Hence, as already noticed in \cite{Binetruy:2014ela}, vacuum energy dictates the position in $R$ of the apparent/trapping horizon, while matter gives its variation in $t$.

\end{document}